\documentclass[graybox]{svmult}

\usepackage{mathptmx}       
\usepackage{helvet}         
\usepackage{courier}        
\usepackage{type1cm}        
\usepackage{makeidx}         
\usepackage{graphicx}        
\usepackage{multicol}        
\usepackage[bottom]{footmisc}

\makeindex             

\begin{document}

\title*{Music and Astronomy}
\titlerunning{Music and Astronomy}
\author{Jos\'e~A. Caballero, Sara Gonz\'alez S\'anchez, Iv\'an Caballero}
\institute{Jos\'e~A. Caballero \at Departamento de Astrof\'{\i}sica y
Ciencias de la Atm\'osfera, Facultad de F\'{\i}sica, Universidad Complutense de
Madrid, E-28040 Madrid, Spain,
\email{caballero@astrax.fis.ucm.es}}
\maketitle

\abstract*{What do Brian May (the Queen's lead guitarist), William Herschel and
the Jupiter Symphony have in common? And a white dwarf, a piano and
Lagartija Nick? 
At first glance, there is no connection between them, nor between Music and
Astronomy.
However, there are many revealing examples of musical Astronomy and
astronomical Music.
This four-page proceeding describes the {\em sonorous} poster that we showed
during the VIII Scientific Meeting of the Spanish Astronomical Society.}

\abstract{What do Brian May (the Queen's lead guitarist), William Herschel and
the Jupiter Symphony have in common? And a white dwarf, a piano and
Lagartija Nick? 
At first glance, there is no connection between them, nor between Music and
Astronomy.
However, there are many revealing examples of musical Astronomy and
astronomical Music.
This four-page proceeding describes the {\em sonorous} poster that we showed
during the VIII Scientific Meeting of the Spanish Astronomical Society.}

\section{\em La M\'usica y la Astronom\'{\i}a}
\label{sec:1}

Music and Astronomy, together with Arithmetic and Geometry, were the
four subjects of the {\em quadrivium} taught in Scholastic medieval
universities: 
``Music is the Continuous In Motion, Astronomy is the Discrete In Motion''.
In Greek mythology, two of the nine Muses were Euterpe, the goddess of Music and
lyric poetry, and Urania, the goddess of Astronomy.
The ibis-headed Egyptian deity Thoth, the Pacifier of the Gods, was the god
of the Music and Astronomy, apart from the Moon, Geometry, Medicine, drawing,
writing...
However, there is no apparent link between Music and Astronomy.

\begin{figure*}[t]
\sidecaption[t]
\includegraphics[width=0.64\textwidth]{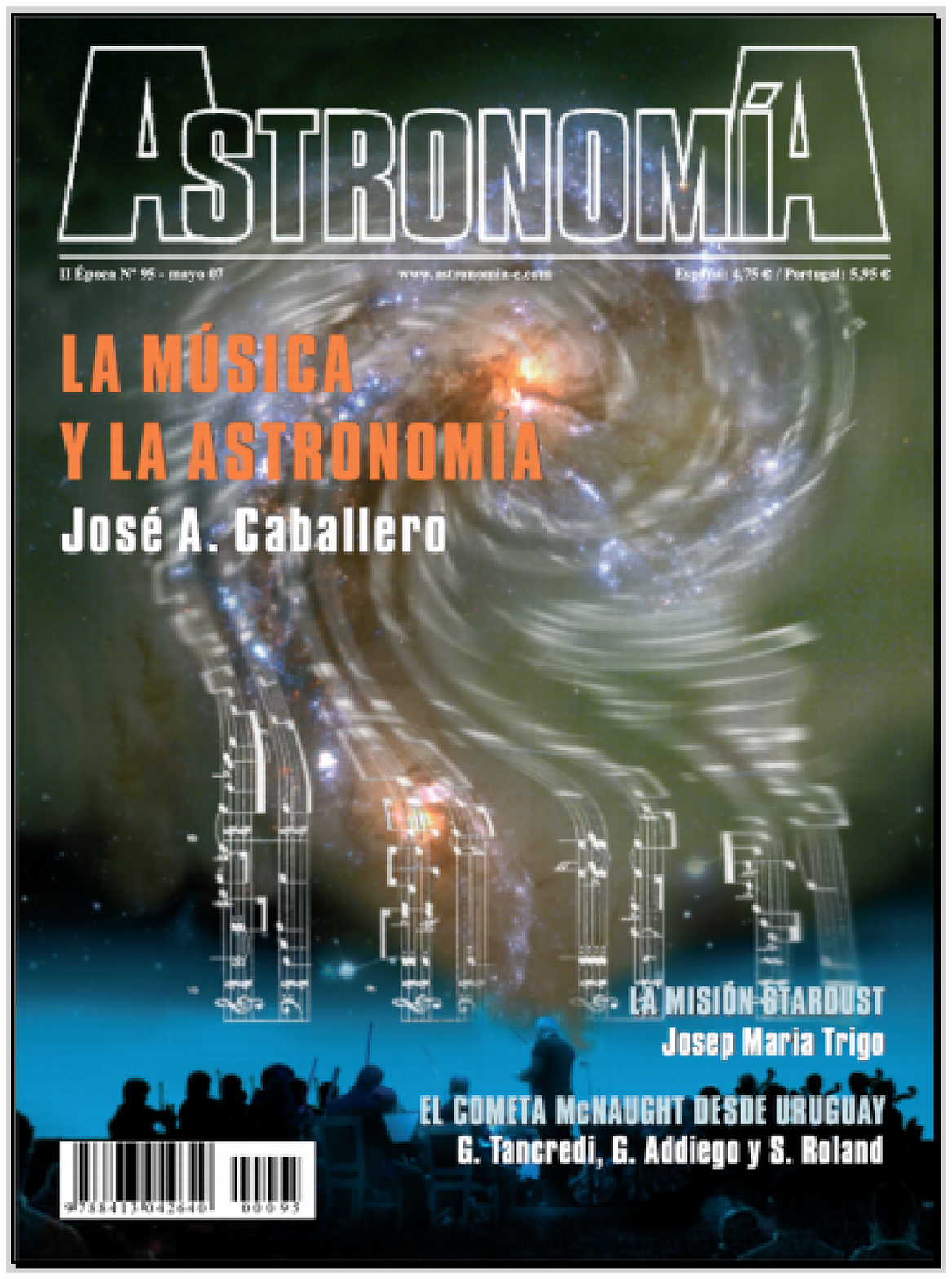}
\caption{Cover page of volume 95 of Astronom\'{\i}A, an international journal on
popular Astronomy in Spanish.
There, Caballero (2007) gave numerous examples of musical Astronomy and
astronomical Music: 
Sir F.~W. Herschel, who discovered Uranus and the infrared radiation after
having composed 24 symphonies and many concertos;
B.~H. May, who became famous with Queen after having worked as an
astronomer at the Observatorio del Teide; 
lyrics of many themes, such as {\em Would you like to look through my
telescope? / The Milky Way's a fine sight to see. / All around the Universe, we
try so hard to view / what's new} (Saved by a bell, Discovery, Mike Oldfield);
CD covers with clear astronomical influence (e.g. PSR~B1919+20 in Joy Division's
Unknown pleasures or Engraved Hourglass Nebula in Pearl Jam's Binaural); etc.}
\label{thecoverpage}       
\end{figure*}

The separation between Music and Astronomy is only superficial, as already
noticed by Pythagoras in his {\em Musica Universalis}.
Sometimes, we (the astronomers) use Music for ``visualising'' some
astrophyisical mechanisms, such as pulsations of white dwarfs and red giants,
magnetic fields and stellar winds of massive stars or energetic phenomena in
outer atmospheres of the Solar System planets.
Caballero (2007) gave numerous examples of musical
Astronomy and astronomical Music (Fig.~\ref{thecoverpage}).
A sequel, {\em La M\'usica y la Astronom\'{\i}a~2}, is in preparation.

\section{``Astromusic''} 
\label{sec:2}

% from we have extracted some text pieces. 

\begin{figure*}
\sidecaption
\includegraphics[width=0.95\textwidth]{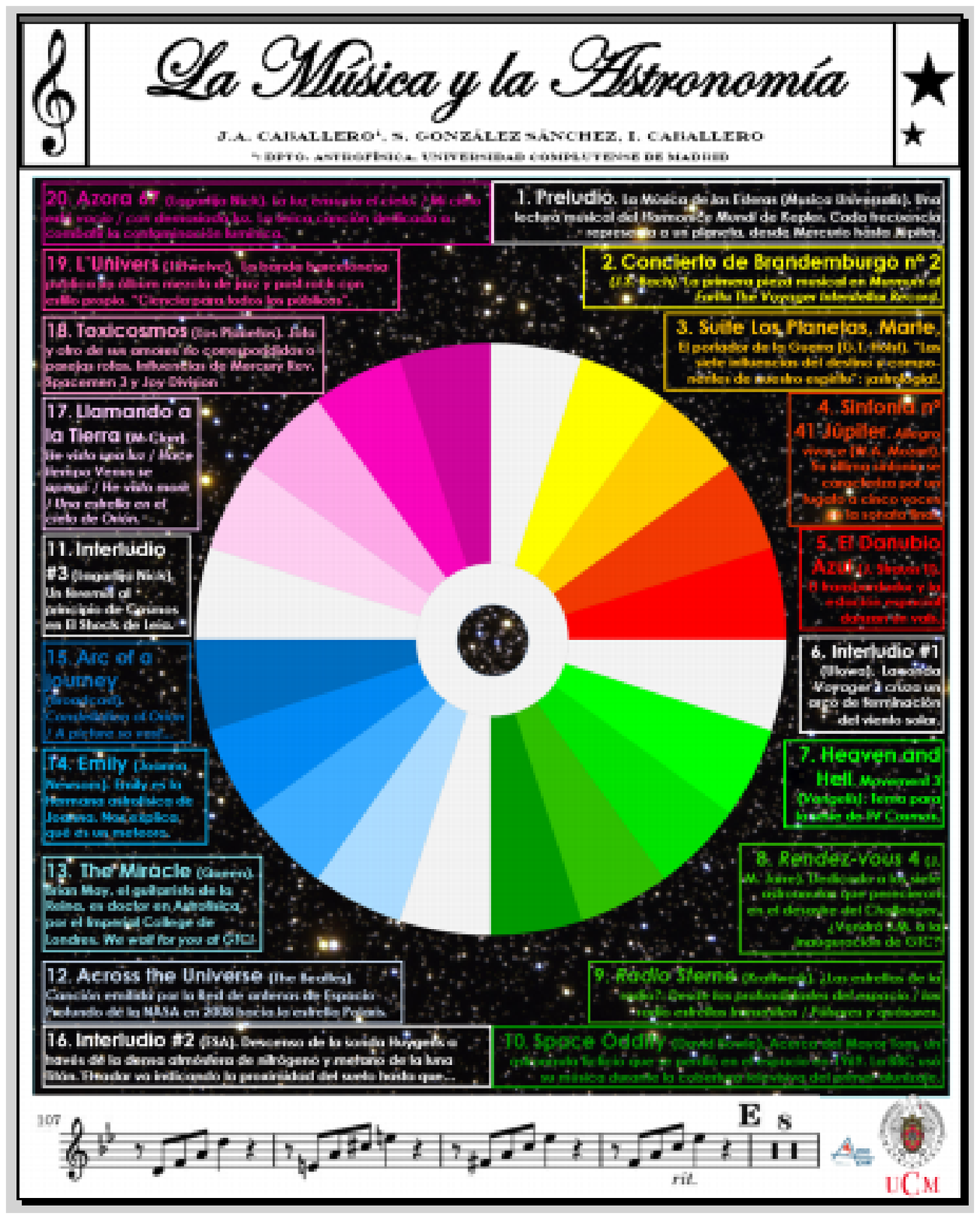}
\caption{Poster presented by the authors in the VIII Scientific Meeting of the
Spanish Astronomical Society. 
A CD player with headphones and speakers was available in front of the poster,
and astronomers could choose any track for listening.} 
\label{theposter}       
\end{figure*}

We prepared a poster (Fig.~\ref{theposter}) and a music compact disc (CD) to be
heard {\em only} during the coffee breaks of the VIII Scientific Meeting of the
Spanish Astronomical Society (Sociedad Espa\~nola de Astronom\'{\i}a) held in
Santander, 7--11 July 2008. 
The contribution fitted the section ``Teaching, dissemination and
popularisation of Astronomy'' (``Ense\~nanza, difusi\'on y divulgaci\'on de la
Astronom\'{\i}a'').

The CD consisted of 20 tracks, splitted into one prelude (track \#1),
three interludes (tracks \#6, \#11 and \#16) and four groups of musical themes
and excerpts separated by the interludes. 
These groups were classical music (tracks \#2--5), electronica and new age
(tracks \#7--9), international pop-rock (tracks \#10 and \#12--15), and Spanish
music (including jazz, pop and rock; tracks \#17--20).
The last track also worked as the postlude of the CD.
The 20 tracks are described next:

\begin{enumerate}
\item[\#1] {\bf Prelude} (Spiegel, Rogers \& Ruff).
The Music of the Spheres ({\em Musica Universalis}), a 20th-century musical
readout of J.~Kepler's {\em Harmonices Mundi}. 
Each frequency represented a planet, from Mercury to Jupiter.
\item[\#2] {\bf Allegro / Brandenburg Concerto No.~2 in F major, BWV 1047}
(J.~S. Bach). 
The first musical piece in {\em Murmurs of Earth: The Voyager Interstellar
Record}. 
\item[\#3] {\bf Mars, the Bringer of War / The Planets Op.~32} (G.~T. Holst).
A seven-movement orchestral suite with an astrological (non-astronomical)
concept. 
%the planets are ``the seven influences of destiny and components of our spirit''.
\item[\#4] {\bf Molto allegro / Symphony No.~41 in C major (K.~551), {\em
Jupiter}} (W.~A. Mozart).    
The last movement of his last {\em sinfonia}, with a remarkable five-voice {\em
fugato}. 
\item[\#5] {\bf On the Beautiful Blue Danube} (J. Strauss II).
The Pan Am space clipper and the Space Station 5 dance a waltz in Stanley
Kubrick's film {\em 2001: A Space Odyssey}. 
\item[\#6] {\bf First interlude} (Gurnett et~al.).
Plasma oscillations when the Voyager~1 interplanetary spacecraft approached the
solar wind termination shock. 
\item[\#7] {\bf Movement 3 / Heaven and Hell (Theme from Cosmos)} (Vangelis).
Main theme of the television documentary series {\em Cosmos}.
Listen also Albedo~0.39, Pulstar...
\item[\#8] {\bf Fourth Rendez-Vous / Rendez-Vous} (J.~M. Jarre).
The last piece of the album was dedicated to the seven astronauts that died in
the {\em Challenger} disaster. 
%UNESCO Goodwill Ambassador
\item[\#9] {\bf Radio Sterne / Radio-Aktivit\"at} (Kraftwerk).
Did video kill the radio stars? {\em Aus des Weltalls Ferne / Funken
Radiosterne / Pulsare und Quasare}.
%(From the deeps of space / Radio stars are transmitting / Pulsars and quasars).
\item[\#10] {\bf Space Oddity / Space Oddity} (D. Bowie).
%``Major Tom's departure from Earth is successful and everything goes according
%to plan, but then he loses contact with Ground Control''.
Released to coincide with the first moon landing, the song tells the story of
the astronaut Major Tom.
\item[\#11] {\bf Second interlude} (ESA).
Radar echoes from Titan's surface:
as the Huygens probe approaches the ground, both the pitch and intensity
increase.
\item[\#12] {\bf Across the Universe / Let it be} (The Beatles).
The song was emitted in 2008 towards the star Polaris by the NASA 70\,m antenna
in Robledo de Chavela.
\item[\#13] {\bf The Miracle / The Miracle} (Queen).
In May 2008, Brian May got his Ph.D. degree with the thesis {\em A Survey of
Radial Velocities in the Zodiacal Dust Cloud}.
We wait for you at the {\em Concierto de las Estrellas} at the Gran
Telescopio Canarias! 
%The Miracle song (Fig.~\ref{thecoverpage}) is one of Brian May's favourites.
%Because of the recent PhD doctorate in Astrophysics of Brian May
%(Fig.~\ref{thecoverpage}), the name of the latest album by Queen + Paul Rodgers
%is The Cosmos Rocks (illustrated with the Eagle Nebula). 
\item[\#14] {\bf Emily / Ys} (Joanna Newsom).
Emily is an astrophysicist, apart from Joanna's sister.
They explain us what the meteorite, meteor and meteoroids are.
\item[\#15] {\bf Arc of a jouney / Tender Buttons} (Broadcast).
{\em Constellation of Orion / A picture with a past / A future so vast}.
Retro-futuristic sci-fi sounds.
\item[\#16] {\bf Third interlude} (Lagartija Nick).
A theremin at the beginning of Cosmos, in El shock de Leia.
Look also for the SETI 1st Theremin Concert for Aliens.
\item[\#17] {\bf Llamando a la Tierra / Usar y tirar} (M-Clan).
{\em He visto una luz / Hace tiempo Venus se apag\'o / He visto morir / una
estrella en el cielo de Ori\'on}.
\item[\#18] {\bf Toxicosmos / Una semana en el motor de un autob\'us} (Los
Planetas). 
Their last album is La leyenda del espacio, a tribute to Camar\'on's
La leyenda del tiempo.
\item[\#19] {\bf L'Univers / L'Univers} (12twelve).
A Barcelonese band with mixes of free-jazz, post-rock and ``Ciencia para todos los
p\'ublicos'' (popular science).
\item[\#20] {\bf Azora 67 / Lagartija Nick} (Lagartija Nick).
A song to fight the light pollution! {\em Demasiada luz / Demasiada luz / Mi
cielo est\'a vac\'{\i}o con demasiada luz}.
\end{enumerate}

\begin{acknowledgement}
JAC thanks Antonio Arias and countless members of the the Sociedad
Espa\~nola de Astronom\'{\i}a for their back up and feed back.
Partial financial support was provided by the Spanish Ministerio Educaci\'on y
Ciencia, the Comunidad Aut\'onoma de Madrid, the Universidad Complutense de
Madrid and the European Social Fund.  
\end{acknowledgement}


\begin{thebibliography}{99.}%

\bibitem{Ca07}
J.~A. Caballero, 
\emph{Astronom\'{\i}A} \textbf{95}, 26 (2007).

\end{thebibliography}
\end{document}